\documentclass[aps,amsmath,amsfonts,11pt]{revtex4}
\usepackage{graphicx}
\usepackage{epstopdf}
\usepackage{epsfig,graphicx}
\usepackage[english]{babel}
\usepackage{amsfonts}
\usepackage{amsmath}
\usepackage{latexsym}
\usepackage{graphics,bm}
\usepackage{dcolumn}
\usepackage{bm}
\usepackage{rotating}

\begin{document}

\title{Spin squeezing and Quantum Fisher Information in the Jaynes-Cummings Dicke Model}

\author{Aranya B Bhattacherjee and Deepti Sharma}

\address{School of Physical Sciences, Jawaharlal Nehru University, New Delhi-110067, India}

\begin{abstract}
We investigate spin squeezing (SS) and the quantum Fisher information (QFI) for the Jaynes-Cummings Dicke (JC-Dicke) model in a two component atomic Bose-Einstein condensate (BEC) inside an optical cavity. Analytical expressions for spin squeezing and the reciprocal of the quantum Fisher information per particle (RMQFI) are derived using the frozen spin approximation. It is shown that in the superradiant phase near the critical point, maximum squeezing and maximum quantum entanglement occurs. The present study is relevant to quantum information processing and precision spectroscopy.

\end{abstract}

\maketitle

\section{Introduction}
Atomic spin squeezing \citep{gross,ma} of a large ensemble of spin systems has potential applications in precision measurements \citep{wine1,wine2,polzik,cronin} and particle entanglement \citep{sorensen,bigelow,gue}. The measurement accuracy of atomic clocks for time and frequency metrology is limited by the quantum mechanical uncertainity of spin operators \citep{wine1,wine2}. Kitagawa and Ueda \citep{kita1} proposed the generation of spin-squeezed states, which redistributes the uncertainity unevenly between the two components of the total angular momentum and the component with the reduced uncertainity can be utilized for precision measurements with applications in atomic clocks \citep{wine1,wine2,tur,meyer,leib} and quantum information processing \citep{sorensen,wang,kor,yi}.Atomic spin squeezing requires nonlinear interactions between spins which can be realized by one-axis twisting and two-axis counter-twisting \citep{kita1}.  Quantum entanglement/correlations among the spins is responsible for spin squeezing. In this context it has been found in various models that quantum Fisher information (QFI) together with spin squeezing(SS) are connected to quantum entanglement \citep{hu,jiang,li}.

In the context of Bose-Einstein condensates, spin-squeezed states can be experimentally achieved \citep{orz,ried,est,mue}.  The Dicke model describes a large number of two-level atoms interacting with a single optical mode. Above a certain critical light-matter coupling strength, leads to a quantum phase transition from a normal phase to a superradiant phase \citep{emary,bren}. The Dicke model has been realized experimentally in a BEC confined in an optical cavity \citep{bren}. In this work, we study the spin squeezing and quantum Fisher information of the Jaynes-Cummings Dicke model realizable in a two-componet atomic BEC in a cavity, which has twisting and coupling in two angular momentum components \citep{li2,li3}.  Approximate analytical expressions of spin squeezing and quantum Fisher information are derived under the frozen spin approximation \citep{law1,bhatt1}. We show for the first time that both maximum SS and quantum entanglement exists in the superradiant phase near the critical point. Spin squeezing dynamics and chaos of the Dicke model has been studied earlier \citep{song} while a recent study explored the influence of Rabi oscillation energy on quantum Fisher information and phase sensitivity of BEC in a double-well potential \citep{zhu}. Exact numerical analysis of spin squeezing in different atomic systems has been performed earlier \citep{jin1,jin2,jin3,chen}.

\section{Model}

We consider a Rb atomic BEC optically trapped in an optical cavity as described in ref. \citep{li2}. Under the two-mode approximation, only the atomic states, $5^{2}{S}_{1/2}$ (${F}=1, {m_{F}}=1$) ground state $|1\rangle$ and ${F}=1, {m_{F}}=0$, metastable state $|2 \rangle $ are considered in this model. These two states $|1\rangle$ and $|2\rangle$ are coupled by an ancillary excited state $5^{2} P_{3/2}$ ($|3\rangle$), where $|1 \rangle \leftrightarrow |3 \rangle$ and  $|2 \rangle \leftrightarrow |3 \rangle$ states coupled by the quantized optical cavity field strength $g_{13}$ and an external classical field strength $\Omega_{32}$ respectively. An effective two-level configuration can be obtained after adiabatically eliminating the ancillary excited state $| 3 \rangle$ under the assumption of large one-photon detuning. Following the work of \citep{li2,li3}, one obtains an effective JC-Dicke Hamiltonian for $N$ number of atoms

\begin{equation}
H=\omega_{0} a^{\dagger} a+ \omega_{a} J_{z}+\frac{\eta}{N}J_{z}^{2}+\frac{\lambda}{\sqrt{N}}a J_{+}+\frac{\lambda^{*}}{\sqrt{N}}a^{\dagger}J_{-},
\end{equation}

where

$\omega_{0}=\omega-\omega_{cl}$, $\eta= \left(  \frac{\eta_{1}+\eta_{2}}{2}-\eta_{12}   \right)$, $\lambda= \lambda_{eff} \sqrt{N}$, $\lambda_{eff}=g_{13} \Omega_{32}/\Delta$ and $\Delta= \omega_{3}-(\nu_{2}+\omega_{2})-\omega_{cl}$ $ >>$  $g_{13},\Omega_{32}$. $\omega_{i}, (i=1,2,3)$ is the internal energy levels for atomic state $|i \rangle$, $\nu_{l}= \int d^{3} \vec{r} [-\nabla^{2}/2m+V(\vec{r}) ] \phi_{l}(\vec{r})$ ($l=1,2$) is the trapped frequency for the states $|1 \rangle$ and $|2 \rangle$ with $V(\vec{r})$ being the trapped potential, m is the atomic mass and $\phi_{l}(\vec{r})$ is the corresponding BEC wavefunction.
The cavity mode annihilation (creation) operator is defined as $a$($a^{\dagger}$) with frequency $\omega$ and $\lambda_{eff}=g_{13}\Omega_{32}/\Delta$ is the reduced effective coupling strength for the two-photon Raman process. Here $\Delta=\omega_{3}-(\nu_{2}+\omega_{2}-\omega_{cl})>>g_{13},\Omega_{32}$ is the large single-photon detuning. Also $\omega_{cl}$ is the frequency of the classical optical field.
$\eta_{l}=(4 \pi a_{l}/m)\int d^{3}\vec{r} |\phi_{l}(\vec{r})|^{4}$ and $\eta_{12}= (4 \pi a_{12}/m) \int d^{3}\vec{r} |\phi_{1}^{*}(\vec{r}) \phi_{2}(\vec{r})|^{2}$ with $a_{l}$ and $a_{12}=a_{21}$, the intraspecies and interspecies $s-$wave scattering lengths, respectively.

Now taking $\lambda=\chi e^{i \phi}$, we rewrite the Hamiltonian of eqn.(1) as

\begin{equation}
H=\omega_{0}a^{\dagger}a+\omega_{a} J_{z}+\frac{\eta J_{z}^{2}}{N}+\frac{\chi}{\sqrt{N}}(a+a^{\dagger})\left(\cos{\phi} J_{x}-\sin{\phi}J_{y} \right)
\end{equation}

In the bad cavity limit, the decay rate of cavity photons ($\kappa$) is much larger than any other rates of the system and the cavity field always follows the atomic dynamics adiabatically. Under this assumption, we can adiabatically eliminate the cavity mode and obtain a Hamiltonian for the atomic degrees of freedom alone.

\begin{equation}
 H = \omega_{a} J_{z} + \frac{\eta J_{z}^{2}}{N}-\left[\Omega_{x} J_{x}^{2}+\Omega_{y} J_{y}^{2}-\sqrt{\Omega_{x} \Omega_{y}} (J_{x} J_{y}+ J_{y} J_{x}) \right],
\end{equation}

\begin{equation}
 \Omega_{x} = \frac{\chi^{2} \omega_{0} \cos^{2}{\phi}}{N(\omega_{0}^{2}+\kappa^{2})}=\Omega_{0} \cos^{2}{\phi},
\end{equation}

\begin{equation}
 \Omega_{x} = \frac{\chi^{2} \omega_{0} \sin^{2}{\phi}}{N(\omega_{0}^{2}+\kappa^{2})}=\Omega_{0} \sin^{2}{\phi},
\end{equation}

\section{Spin Squeezing and Quantum Fisher Information}

In this section, we will work in the Heisenberg picture and study in detail the spin squeezing and quantum Fisher information in the Hamiltonian of eqn.(3). In particular, we will explore the dependence of SS and QFI on the atom-photon coupling strength.

\subsection{Definitions}

Following the criteria of Kitagawa and Ueda of spin squeezing \citep{kita1}, we introduce the squeezing parameter

\begin{equation}
 \xi^{2}=\frac{4 (\Delta J_{\theta})^{2}_{min}}{N},
\end{equation}

where $(\Delta J_{\theta})^{2}_{min}$ is the minimal spin fluctuation in a plane perpendicular to the mean spin direction. When $\xi^{2} \geq 1$, the spin systems are uncorrelated while $\xi^{2}<1$ signifies spin squeezing and a signature of entanglement.

Now in order to study QFI in our system, we follow \citep{pezze,liu1} who in order to characterize multi-particle entanglement in an $N$ qubit state had introduced the following quantity $\beta$ (termed as RMQFI),

\begin{equation}
 \beta^{2} = \frac{N}{F_{Q}[\rho_{in}, J_{\vec{n}}]},
\end{equation}

where, $F_{Q}[\rho_{in}, J_{\vec{n}}]$ is the quantum Fisher information and $\vec{n}$ is an arbitrary direction. Here, $\beta^{2}<1$ implies multi-particle entanglement. Following \citep{hu,jiang,li}, the expression for $\beta^{2}$ can be written for our system as,

\begin{equation}
 \beta^{2}=\frac{N}{4 (\Delta J_{\theta})^{2}_{max}},
\end{equation}

where, $(\Delta J_{\theta})^{2}_{max}$ is the maximum variance in a plane perpendicular to the mean spin direction.

 \subsection{Theoretical Results}

The Hamiltonian (1) demonstrates a second-order superradiant quantum phase transition at the critical point $\chi_{c}$. In order to determine the critical point $\chi_{c}$, we start with the Heisenberg equations of moton of the angular momentum operators $J_{x}$, $J_{y}$, $J_{z}$ and the steady state values of the cavity field operators $(a+a^{\dagger})$.

\begin{equation}
  \dot{J_{x}}(t)= - \omega_{a} J_{y}(t)-\frac{\eta}{N} (J_{z}(t) J_{y}(t)+ J_{y}(t) J_{z}(t))-  \frac{\chi}{\sqrt{N}} (a+a^{\dagger}) \sin{\phi} J_{z}(t),
\end{equation}

\begin{equation}
  \dot{J_{y}}(t)=  \omega_{a} J_{x}(t)+ \frac{\eta}{N} (J_{z}(t) J_{x}(t)+ J_{x}(t) J_{z}(t))- \frac{\chi}{\sqrt{N}} (a+a^{\dagger}) \cos{\phi} J_{z}(t),
\end{equation}

\begin{equation}
 \dot{J_{z}}(t)= \frac{\chi}{\sqrt{N}} (a+a^{\dagger})(J_{y}(t) \cos{\phi}+J_{x}(t) \sin{\phi}),
\end{equation}

\begin{equation}
 (a+a^{\dagger})= -\frac{2 \chi \omega_{0}}{\sqrt{N} (\omega_{0}^{2}+\kappa^{2})}\left( \cos{\phi} J_{x}- \sin{\phi} J_{y} \right).
\end{equation}

A steady state analysis for $\left\langle J_{z} \right\rangle = -N/2$ (corresponds to the normal phase i.e. all atoms are in the lower state) of the above mean field equations of motion (for large $N$) yields the critical value as

\begin{equation}
 \chi_{c}= \sqrt{\frac{(\omega_{0}^{2}+\kappa^{2})(\omega_{a}-\eta)}{\omega_{0}}}.
\end{equation}

\bigskip

Starting from the Hamiltonian (3), we now examine the generation and coherent control of spin squeezing and quantum Fisher information. We prepare the system to start from the lowest eigenstate of $J_{z}$, $J_{z}|J,-J  \rangle_{z} = -J |J,-J \rangle_{z} $. In order to investigate the quantum spin squeezing and quantum Fisher information, we examine the Heisenberg equations of motion of $J_{x}(t)$ and $J_{y}(t)$ derived from Hamiltonian (3),

\begin{equation}
 \dot{J}_{x}(t)=-\omega_{a} J_{y}(t)-\left( \frac{\eta}{N}+\Omega_{y}\right)\left(J_{z}(t) J_{y}(t)+J_{y}(t) J_{z}(t) \right)+\sqrt{\Omega_{x} \Omega_{y}} \left( J_{x}(t) J_{z}(t) + J_{z}(t) J_{x}(t) \right),
\end{equation}

\begin{equation}
 \dot{J}_{y}(t)=\omega_{a} J_{x}(t)-\left( \frac{\eta}{N}+\Omega_{x}\right)\left(J_{z}(t) J_{x}(t)+J_{x}(t) J_{z}(t) \right)+\sqrt{\Omega_{x} \Omega_{y}} \left( J_{y}(t) J_{z}(t)+J_{z}(t) J_{y}(t) \right),
\end{equation}

If $\Omega_{x} (\Omega_{y}) >> \eta/N$, $\omega_{a}$, then the external field forced the total spin to freeze along the negative $z-$ direction i.e. $\langle J_{z}(t) \rangle = -J$. This approximation is termed as frozen spin approximation \citep{law1,bhatt1} which permits harmonic solutions of $J_{x}(t)$ and $J_{y}(t)$,

\begin{equation}
 J_{x}(t)=\left\lbrace \cos{\omega t}-\frac{\Omega_{2}}{\omega} \sin{\omega t} \right\rbrace J_{x}(0)+\frac{\Omega_{1}}{\omega} \sin{\omega t} J_{y}(0),
\end{equation}

\begin{equation}
 J_{y}(t)=\left\lbrace \cos{\omega t}-\frac{\Omega_{2}}{\omega} \sin{\omega t} \right\rbrace J_{y}(0)+\frac{\Omega_{1}}{\omega} \sin{\omega t} J_{x}(0),
\end{equation}

where,

\begin{equation}
 \Omega_{1}=2 J \left( \frac{\eta}{N}+\Omega_{y}\right)-\omega_{a}, \nonumber
\end{equation}

\begin{equation}
 \Omega_{2}= 2 J \sqrt{\Omega_{x} \Omega_{y}}, \nonumber
\end{equation}

\begin{equation}
 \Omega_{3}= \omega_{a}- 2 J \left( \frac{\eta}{N}+ \Omega_{x}\right), \nonumber
\end{equation}

\begin{equation}
 \omega = \sqrt{\Omega_{1} \Omega_{3}+\Omega_{2}^{2}}= (\omega_{a}-\eta) \sqrt{\left(\frac{\chi^{2}}{\chi_{c}^{2}}-1 \right) }.
\end{equation}

As evident from the above expressions, the analysis is valid in the superradiant phase i.e $\chi > \chi_{c}$, since for $\chi<\chi_{c}$, $\omega$ is not real.  When the system starts from $|J, m_{z}=-J \rangle$, the only non-vanishing spin component is $J_{z}(t)$ because $\langle J_{x}(t)\rangle$ $=$ $\langle J_{y}(t) \rangle$ $=$ $0$ at all times. Now with the help of $J_{x}(t)$ and $J_{y}(t)$, we can investigate quantum spin squeezing and quantum Fisher information.
If $\langle J_{x}(t) J_{y}(t)\rangle$ $=$ $\langle J_{y}(t) J_{x}(t)\rangle$ $= 0$, i.e. there is no correlation between $J_{x}(t)$ and $J_{y}(t)$, the increase or reduced spin fluctuations occur either in the $x$ or $y$ direction. However in our case,

\begin{equation}
 \langle J_{x}(t) J_{y}(t) \rangle + \langle J_{y}(t) J_{x}(t) \rangle = \frac{J}{2} \left\lbrace \frac{(\Omega_{1}+\Omega_{3})}{\omega} \sin{2 \omega t}+2 \frac{\Omega_{2}}{\omega^{2}}(\Omega_{1}-\Omega_{3}) \sin^{2}{\omega t} \right\rbrace.
\end{equation}

Consequently since $\langle J_{x}(t) J_{y}t)\rangle$+$\langle J_{y}(t) J_{x}(t)\rangle$ $\neq$ $0$, the reduced spin fluctuations neither occurs along $x$ nor along $y$ direction.

Now in order to determine the increased and reduced spin fluctuation directions, we introduce the spin component,

\begin{equation}
 J_{\theta}(t)=\vec{J}(t). \vec{n}_{\theta}= J_{x}(t) \cos{\theta}+J_{y}(t) \sin{\theta},
\end{equation}

where the unit vector $\vec{n}_{\theta}= x \cos{\theta}+y \sin{\theta}$, with $\theta$ as the angle between $x$ axis and $\vec{n}_{\theta}$. Since $\langle J_{x}(t)\rangle$ $=$ $\langle J_{y}(t)\rangle$ $=$ $0$, the fluctuation of the spin component $J_{\theta}(t)$ reads,

\begin{equation}
 (\Delta J_{\theta}(t))^{2}=\frac{1}{2} \left\lbrace \langle J_{x}^{2}(t)+J_{y}^{2}(t)\rangle  + \langle J_{x}^{2}(t)-J_{y}^{2}(t)\rangle \cos{2 \theta} \right\rbrace + \frac{1}{2} \langle J_{x}(t) J_{y}(t) +J_{y}(t) J_{x}(t)\rangle \sin{2 \theta}.
\end{equation}

The optimally squeezed angle $\theta_{opt}$ is obtained via minimizing $(\Delta J_{\theta})^{2}$ with respect to $\theta$. This yields,

\begin{equation}
 (\Delta J_{\theta}(t))^{2}_{min(max)}= \frac{1}{2} \left\lbrace \langle J_{x}^{2}+J_{y}^{2} \rangle \mp \sqrt{\langle J_{x}^{2}-J_{y}^{2}\rangle+\langle J_{x}J_{y}+J_{y} J_{x} \rangle ^{2}} \right\rbrace.
\end{equation}.

The squeezing parameter $\xi$ and RMQFI $\beta$ are determined as,

\begin{equation}
 \xi=\left[ \frac{\frac{\chi^{2}}{\chi_{c}^{2}} \left( \frac{\chi^{2}}{\chi_{c}^{2}} -2\right)+2  }{\frac{\chi^{2}}{\chi_{c}^{2}}-1} \sin^{2}{\omega t} + 2 \cos{\omega t} -\left\lbrace \frac{\frac{\chi^{4}}{\chi_{c}^{4}} \left( \frac{\chi^{2}}{\chi_{c}^{2}} -2\right)^{2}}{(\frac{\chi^{2}}{\chi_{c}^{2}}-1)^{2}} \sin^{4}{\omega t}+ \frac{\frac{\chi^{4}}{\chi_{c}^{4}}}{\frac{\chi^{2}}{\chi_{c}^{2}}-1} \sin^{2}{2 \omega t} \right\rbrace^{1/2} \right]^{1/2},
\end{equation}

\begin{equation}
 \beta=\left[ \frac{\frac{\chi^{2}}{\chi_{c}^{2}} \left( \frac{\chi^{2}}{\chi_{c}^{2}} -2\right)+2  }{\frac{\chi^{2}}{\chi_{c}^{2}}-1} \sin^{2}{\omega t} + 2 \cos{\omega t} +\left\lbrace \frac{\frac{\chi^{4}}{\chi_{c}^{4}} \left( \frac{\chi^{2}}{\chi_{c}^{2}} -2\right)^{2}}{(\frac{\chi^{2}}{\chi_{c}^{2}}-1)^{2}} \sin^{4}{\omega t}+ \frac{\frac{\chi^{4}}{\chi_{c}^{4}}}{\frac{\chi^{2}}{\chi_{c}^{2}}-1} \sin^{2}{2 \omega t} \right\rbrace^{1/2} \right]^{1/2}.
\end{equation}

\begin{figure}[ht]
\hspace{-0.0cm}
\begin{tabular}{cc}
\includegraphics [scale=0.55]{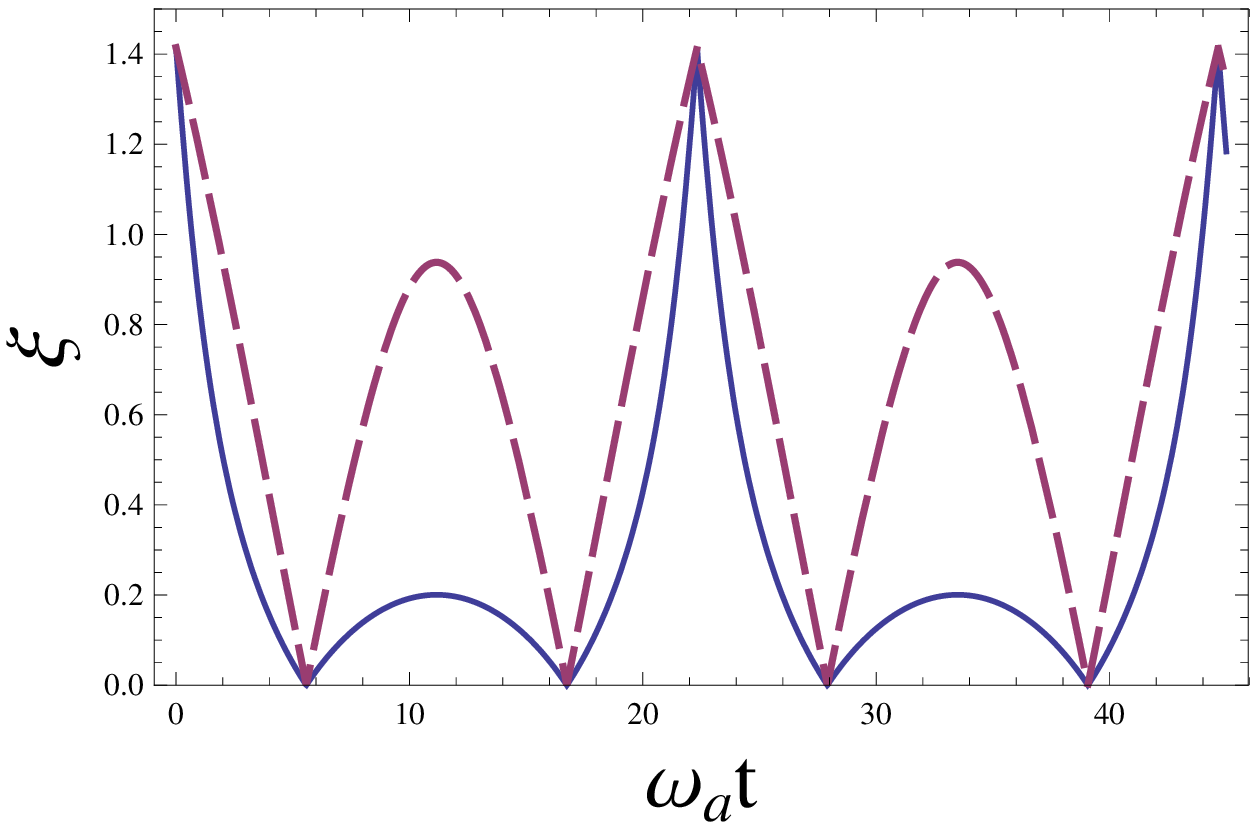} \includegraphics [scale=0.55] {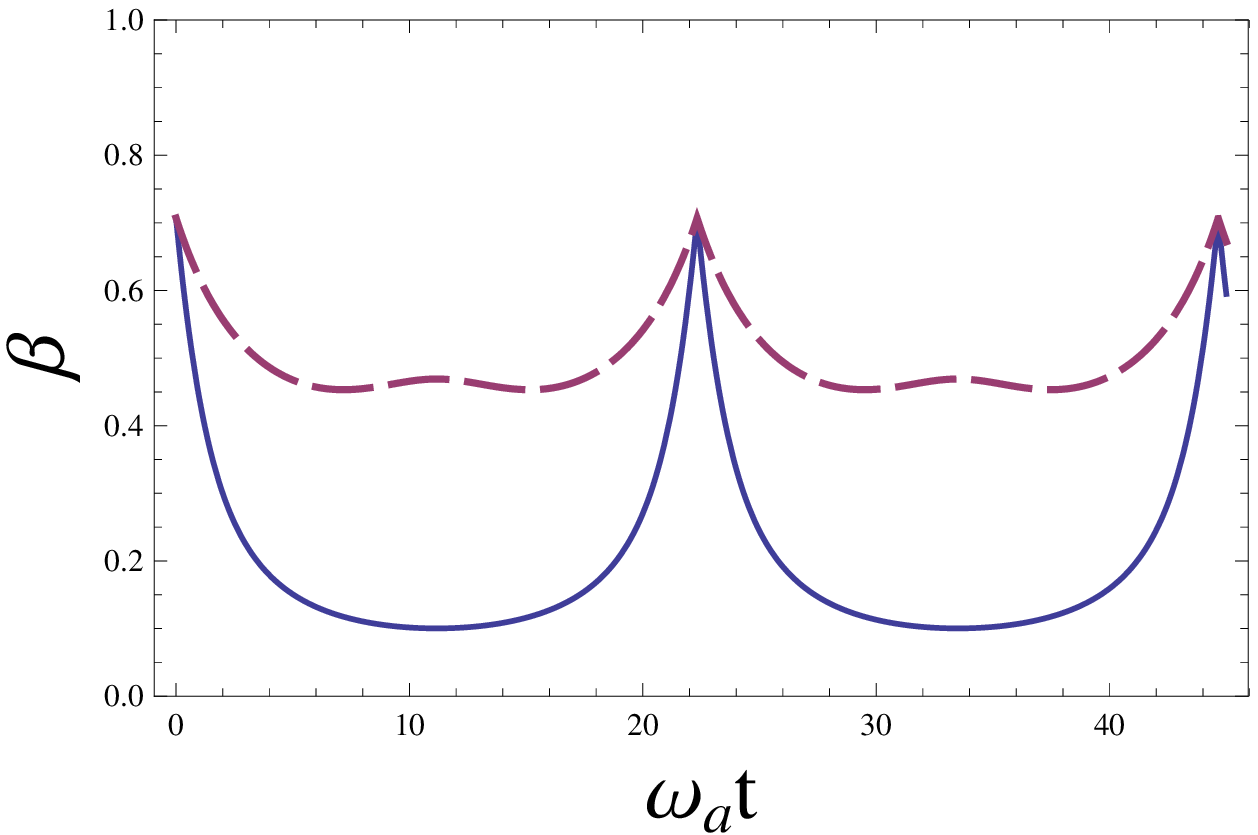}\\
 \end{tabular}
\caption{Spin squeezing parameter (left plot) $\xi$ and RMQFI (right plot) $\beta$ versus $\omega_{a}t$ for  $\chi/\chi_{c}=1.01$ (solid line) and $\chi/\chi_{c}=1.2$(dashed line). Other parameters are $\kappa/\omega_{a}=0.1$, $\omega_{0}/\omega_{a}=1.0$ and $\eta/\omega_{a}=0.01$.}
\label{f1}
\end{figure}

Fig.1 shows the plot of spin squeezing parameter $\xi$ (left plot) and the quantum Fisher information $\beta$ (right plot) as a function of $\omega_{a} t$ for two values of $\chi/\chi_{c}=1.01$ (solid line) and $\chi/\chi_{c}=1.2$(dashed line). Clearly for $\chi/\chi_{c}=1.01$, both $\xi$ and $\beta$ has minimum values compared to that for  $\chi/\chi_{c}=1.2$. This clearly demonstrates that near the critical point in the superradiant phase maximum spin squeezing and maximum quantum entanglement occurs. The close connection between spin squeezing and entanglement has been demonstrated earlier \citep{sorensen,song}. As a function of times, periodically maximum squeezing and entanglement reoccurs. The fig.1 also illustrates that initially at $t=0$, there is no spin squeezing but as the dynamics evolves, spin squeezing occurs. The atom-photon coupling then emerges as coherent handle to control the spin squeezing and quantum entanglement.In a recent experiment, Bell correlations was reported between pseudo spins of about 480 atoms in a spin-squeezed BEC \citep{schmied}. Bell correlations also suggests entanglement. The present work is highly relevant to quantum information processing and precision spectroscopy.

\section{Conclusions}

In conclusion, we have investigated the spin squeezing and quantum Fisher information dynamics in a nonlinear system composed of a Bose-Einstein condensate in an optical cavity. The nonlinearity emerges from the twisting and coupling of the $J_{x}(t)$ and $J_{y}(t)$ components of the angular momentum.The system represents a variant of the Dicke model which undergoes a Dicke quantum phase transition from the normal phase to a superradiant phase above a critical atom-photon coupling strength.It is found that near the critical point in the superradiant phase maximum spin squeezing can be achieved. At the same point where $\xi$ is minimum, $\beta$ is also found to be minimum which signifies that near the critical point in the superradiant phase, maximum quantum entanglement exists.

\begin{acknowledgements}
A. Bhattacherjee acknowledges financial support from the University Grants Commission, New Delhi under the UGC-Faculty Recharge Programme. Deepti Sharma acknowledges financial assistance from Jawaharlal Nehru University.
\end{acknowledgements}

\end{document}